\pdfoutput=1

\documentclass[10pt]{article}

\usepackage{authblk}                
\usepackage[square,numbers]{natbib} 

\usepackage{amsmath,amssymb}                

\usepackage{algorithmic}
\usepackage{graphicx}
\usepackage{textcomp}
\usepackage{xcolor}
\usepackage{url}
\usepackage{listings}

\newcommand\YAMLcolonstyle{\color{red}\mdseries}
\newcommand\YAMLkeystyle{\color{black}\bfseries}
\newcommand\YAMLvaluestyle{\color{blue}\mdseries}

\makeatletter

\newcommand\language@yaml{yaml}

\expandafter\expandafter\expandafter\lstdefinelanguage
\expandafter{\language@yaml}
{
	keywords={true,false,null,y,n},
	keywordstyle=\color{darkgray}\bfseries,
	basicstyle=\YAMLkeystyle,                                 
	sensitive=false,
	comment=[l]{\#},
	morecomment=[s]{/*}{*/},
	commentstyle=\color{purple}\ttfamily,
	stringstyle=\YAMLvaluestyle\ttfamily,
	moredelim=[l][\color{orange}]{\&},
	moredelim=[l][\color{magenta}]{*},
	moredelim=**[il][\YAMLcolonstyle{:}\YAMLvaluestyle]{:},   
	morestring=[b]',
	morestring=[b]",
	literate =    {---}{{\ProcessThreeDashes}}3
	{>}{{\textcolor{red}\textgreater}}1     
	{|}{{\textcolor{red}\textbar}}1 
	{\ -\ }{{\mdseries\ -\ }}3,
}

\lstdefinelanguage{json}{
	basicstyle=\ttfamily\footnotesize,
	showstringspaces=false,
	breaklines=true,
	frame=single,
	captionpos=b,
	literate=
	*{0}{{{\color{blue}0}}}{1}
	{1}{{{\color{blue}1}}}{1}
	{2}{{{\color{blue}2}}}{1}
	{3}{{{\color{blue}3}}}{1}
	{4}{{{\color{blue}4}}}{1}
	{5}{{{\color{blue}5}}}{1}
	{6}{{{\color{blue}6}}}{1}
	{7}{{{\color{blue}7}}}{1}
	{8}{{{\color{blue}8}}}{1}
	{9}{{{\color{blue}9}}}{1}
	{:}{{{\color{black}{:}}}}{1}
	{,}{{{\color{black}{,}}}}{1}
	{\{}{{{\color{black}{\{}}}}{1}
	{\}}{{{\color{black}{\}}}}}{1}
	{[}{{{\color{black}{[}}}}{1}
	{]}{{{\color{black}{]}}}}{1}
	{true}{{{\color{teal}true}}}{4}
	{false}{{{\color{teal}false}}}{5}
	{null}{{{\color{teal}null}}}{4},
	string=[s]{"}{"},
	stringstyle=\color{orange},
	comment=[l]{//},
	commentstyle=\color{gray},
	morecomment=[s]{/*}{*/},
}

\usepackage{xspace}
\newcommand{\tool}{IsingTester\xspace}
\newcommand{\benchmark}{IsingBench\xspace}
\newcommand{\link}{\url{https://github.com/WSE-Lab/IsingBench}}

\usepackage{geometry}
\title{Ising-based Test Optimization and Benchmarking\thanks{This work is funded by the Fundamental Research Funds for the Central Universities and the National Science Foundation of China (No. 62502022).}}

\author{Yige Yang, Man Zhang, Tao Yue}

\affil[]{Beihang University}
\affil[]{\{yangyige, manzhang, yuetao\}.buaa.edu.cn}

\date{}

\begin{document}

\maketitle

\begin{abstract}
Test optimization contains test case selection and minimization, which is an important challenge in software testing and has been addressed with search-based approaches intensively in the past. Inspired by the recent advancement of using quantum optimization solutions for addressing test optimization problems, we looked into Coherent Ising Machines (CIM), which offer potential for solving combinatorial optimization problems, but have not yet been exploited in test optimization. Hence, in this paper, we present \tool, an open-source, Python-based command-line tool that provides an end-to-end pipeline for solving test optimization problems that are formulated as Ising models. 
With \tool, we reformulate test selection and minimization as Ising spin configurations, encode multiple optimization strategies into Ising Hamiltonians, and implement solvers including CIM simulation and brute-force search.
Given a user-provided dataset and solver configuration, \tool automatically performs problem encoding, optimization, and spin decoding, returning selected test cases back to the user. 
Along with \tool, we also present the accompanying \benchmark for evaluating and comparing optimization techniques across Ising-based paradigms against baseline approaches. 
A screencast demonstrating the tool is available at: \link.
\end{abstract}



\section{Introduction}
Non-trivial software projects routinely maintain test suites comprising hundreds and thousands of test cases that are accumulated over years of continuous development. While comprehensive test coverage is essential for software quality, the execution of an entire test suite is increasingly impractical due to time and computational resource limitations. This situation gives rise to test optimization problems, a family of combinatorial optimization challenges, which are typically categoried into Test Case Selection (TCS) and Test Case Minimization (TCM) problems. TCS aims to identify a subset of test cases that retains strong fault-detection capability or other effectiveness objectives while satisfying practical constraints such as execution time and available testing budget. TCM goes a step further, seeking to reduce the overall size of the test suite, while still preserving a satisfactory level of fault coverage or other effectiveness properties such as diversity. Both problems are NP-hard in a general context, and decades of research have produced various techniques to address them, including greedy heuristics-based approaches\cite{miranda2017scope}, clustering-based approaches\cite{coviello2018clustering}, and search-based approaches\cite{zhang2019uncertainty, lu2021search}.

Inspired by the recent advancement of applying quantum optimization solutions (quantum annealing (QA), Quantum Approximate Optimization Algorithm (QAOA)) for addressing classical test optimization problems ~\cite{wang2024quantum, wang2024test}, and a relative longer history of applying quantum-inspired solutions for addressing software engineering optimization problems in general~\cite{zhang2025empirical, zhang2025quantum}, we focus on the potential of applying Coherent Ising Machines (CIM) in solving test optimization problems. This is also motivated by the obsevation that CIM has recently demonstrated promising scalability for solving combinatorial optimization problems by exploiting physical phenomena to escape local optima efficiently~\cite{inui2022control}. CIM natively solves problems formulated as Ising models~\cite{lucas2014ising}: energy-minimization problems over binary spin variables. 

However, we lack a framework that: 1) systematically bridges test optimization problem instances to Ising-model formulations, 2) supports multiple encoding strategies to accommodate heterogeneous project data, and 3) exposes both classical and quantum-inspired solvers through a single, accessible interface. As a result, researchers and practitioners who wish to leverage CIM or similar methods (e.g., QA and simulated annealers (SA) for test optimization face steep barriers: they must manually derive problem encodings, implement solver interfaces, and interpret raw spin outputs. This is a process that is both technically intricate and highly error-prone. 

In this paper, we present \tool, a novel Python tool that addresses this gap by providing a unified, end-to-end pipeline for solving test optimization problems through the Ising model computational paradigm. \tool is, to the best of our knowledge, the very first tool to unify both TCS and TCM under a single Ising-based framework. The tool allows users to select from multiple encoding strategies---each tailored to different types of project data such as execution time and failure rate---and to choose from a set of solvers including CIM and SA\cite{delahaye2018simulated}. Given a problem encoding and solver configuration, \tool automatically translates the test optimization instance into an Ising Hamiltonian, invokes the selected solver to minimize the energy landscape, decodes the resulting spin configuration, and returns the optimized set of test cases to the user. This pipeline abstracts away the mathematical and implementation complexity of Ising-model optimization, making quantum-inspired test optimization accessible to practitioners. We also present \benchmark, which is developed to evaluate and compare optimization techniques across Ising-based paradigms against baseline approaches, and release it at: \url{https://github.com/WSE-Lab/IsingBench}.

\textit{Structure}. Section~\ref{sec:tool} presents \tool, followed by the demonstration of its application (Section~\ref{sec: application}). In Section~\ref{sec:discussions} we discuss the limitations of \tool and how it can be extended. We conclude the paper in Section~\ref{sec:conclusion}.

\section{\tool and \benchmark}\label{sec:tool}

\begin{figure*}
	\centering
	\includegraphics[width=0.8\textwidth]{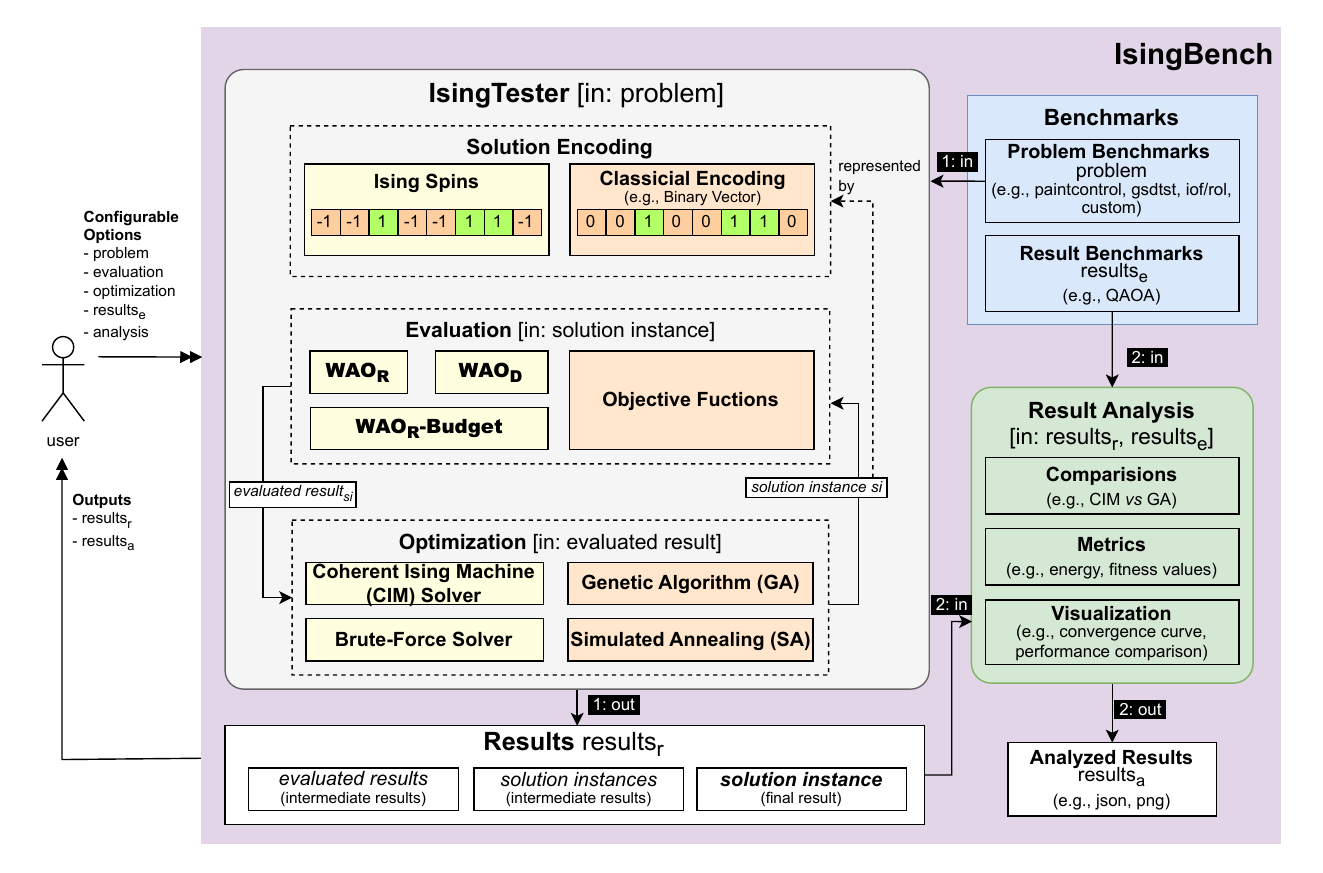}
	\caption{Overview of \tool and \benchmark.}
	\label{fig:structure}
\end{figure*}

Fig.~\ref{fig:structure} presents an overview of \tool and \benchmark. 
\tool is designed to solve test optimization problems using Ising models, while \benchmark is developed to evaluate and compare optimization techniques across Ising-based paradigms against baseline approaches.
Together, these components provide a unified framework for solving software testing optimization problems, comparing different solutions, and analyzing their performance across various benchmarks.

As shown in Fig.~\ref{fig:structure}, the workflow begins with the user configurations of the problem, evaluation strategy, optimization algorithm, CIM configurations and analysis settings. 
The problem can be selected from the predefined benchmarks in \textit{Benchmark} or defined by the user.
Within the \tool module, the problem is first transformed into candidate solution representations using \textit{Solution Encoding}. 
The encoded solutions are then evaluated in \textit{Evaluation} using either existing or customized optimization strategies.
Solutions represented as Ising spins are evaluated using an Ising Hamiltonian (highlighted in yellow), while classically encoded solutions are evaluated using objective functions (highlighted in orange).
To solve the optimization problem, \tool integrates multiple predefined optimization solvers that iteratively improve candidate solutions based on evaluation results (see \textit{Optimization}). 
These solvers include both Ising-based approaches (highlighted in yellow) and classical heuristic algorithms (highlighted in orange). 
The process generates intermediate results, such as intermediate solution instances and their corresponding evaluation values, as well as the final solution instances.

These results are then passed to the \textit{Result Analysis} module for further analysis, etc., computing energy or fitness values and generating visualizations (e.g., convergence curves).
This module not only compares results produced by different optimization approaches within \tool, but also enables comparisons with results from the literature ~\cite{wang2024quantum} when the same problem benchmarks are used.
More details about the modules are presented in the rest of the section. 
Due to space limitations, we mainly focus on the Ising-based optimization.

\subsection{Optimization Strategies (Evaluation)}\label{subsec: optStrategies}

The \textit{Evaluation} module is a core component of \tool, responsible for assessing candidate solutions for a given problem. 
Depending on the selected solution encoding, this module supports two evaluation paradigms. 
When the problem is formulated as an Ising Hamiltonian, candidate solutions are represented as Ising spins, which are evaluated according to their energy values.  
The module evaluates classical candidate solutions using problem-specific objective functions. 
In addition, the module supports to decode Ising spin configurations into corresponding test case selections.

The Ising model is formally represented as Equation~\ref{equation:ising_model}
\begin{equation}
	E(\textbf{s})=-\sum_ih_is_i-\frac{1}{2}\sum_{i,j}J_{ij}s_is_j
	\label{equation:ising_model}
\end{equation}
where $s_i\in \{ -1,+1\}$ are spin variables, $h_i$ are linear coefficients, and $J_{ij}$ are quadratic coefficients. 

\tool currently implements three optimization strategies shown below. 
Two are from the literature, while the third is proposed in this work to demonstrate how our tool can be used to incorporate constraints in Ising-based optimization.

\paragraph{Weighted Attribute Optimization (Ratio-based) (WAO\textsubscript{R})}
This strategy is from Wang et al.'s work on using QAOA for test optimization~\cite{wang2024quantum}. Each spin variable $s_i$ represents the selection decision for test case $i$, where $s_i=-1$ and $s_i=+1$ denote selected and not selected. Each test case carries multiple attributes, each assigned a scalar weight and classified as either an effectiveness measure or a cost one. Optionally, minimization can be included as an additional cost term.  The method is referred to as \textit{ratio-based} because the fitness of each attribute is computed as the ratio of the weighted sum of selected test cases to the total sum across all test cases, as formulated in Equation~\ref{equation:WAO_r_fitness}.
\begin{equation}
	fv=\omega_0(f_0)^2+\dots+\omega_{m-1}(f_{m-1})^2
	\label{equation:WAO_fitness_function}
\end{equation}
\begin{equation}
	f_k(\textbf{s})=\frac{1}{2}(1+\lambda_k\frac{\sum_{i=0}^{n-1}c_i^ks_i}{\sum_{i=0}^{n-1}c_i^k}), s_i\in\{+1,-1\}
	\label{equation:WAO_r_fitness}
\end{equation}
where $c_i^k$ is the $k$th attribute of test case $i$, $\lambda_k=-1$ for cost attributes and $\lambda_k=+1$ for effectiveness attributes. Ising energy $E(\textbf{s})$ is defined as fitness value $fv(\textbf{s})$ in Equation~\ref{equation:WAO_fitness_function}.

\paragraph{Weighted Attribute Optimization (Deviation-based) (WAO\textsubscript{D})}
Adopted from Wang et al.\cite{wang2024test}, this method shares the same input format and configuration as WAO\textsubscript{R}: each test case carries multiple user-weighted effectiveness and cost attributes, with optional minimization as an additional cost, but differs in its fitness function. The method is referred to as \textit{deviation-based} because the fitness of each attribute measures the deviation of the selected subset from a theoretical optimum $L^k$, penalizing configurations that fall short of the ideal value, as formulated in Equation~\ref{equation:WAO_d_fitness}.
\begin{equation}
	f_k(\textbf{s})=(\sum_{i=0}^{n-1}c_{i}^k(\frac{1-s_i}{2})-L^k)^2
	\label{equation:WAO_d_fitness}
\end{equation}
where $L^k$ is the theoretical limit of attribute $k$, the sum of all values for effectiveness attributes, and 0 for cost attributes. As with WAO\textsubscript{R}, Ising energy $E(\textbf{s})$ is defined as the fitness value $fv(\textbf{s})$ in Equation~\ref{equation:WAO_fitness_function}.

\paragraph{Weighted Attribute Optimization with Budget Constraint (Ratio-based) (WAO\textsubscript{R}-Budget)}
This method extends WAO\textsubscript{R} by incorporating a user-defined budget constraint, which specifies the maximum proportion of test cases that should be selected (e.g., a budget of 10 indicates that 10\% of the total test suite should be selected). This formulation demonstrates the capability of \tool to handle constrained optimization problems, where the budget requirement is handled differently depending on the solver type. For Ising solvers, the constraint is integrated directly into the Ising model as a penalty term. The full Ising model is given in Equation~\ref{equation:WAO_r_budget_ising}, where $H_{constraint}$ is defined in Equation~\ref{equation:WAO_r_budget_constraint}
\begin{equation}
	H = fv(\textbf{s}) + H_{constraint}
	\label{equation:WAO_r_budget_ising}
\end{equation}
\begin{equation}
	H_{constraint} = \alpha\left(\sum_{i}^{n-1}\frac{1-s_i}{2} - B\right)^2
	\label{equation:WAO_r_budget_constraint}
\end{equation}
where $B$ is the target number of selected test cases derived from the user-provided budget percentage, and $\alpha$ is a penalty coefficient controlling the strictness of the constraint.

For classical solvers, the budget requirement is instead provided as a separate constraint function, allowing the solvers to handle the objective and constraint independently according to their own optimization strategies. The fitness function follows Equation~\ref{equation:WAO_fitness_function} and Equation~\ref{equation:WAO_r_fitness}, as defined in WAO\textsubscript{R}.

\subsection{Optimization Techniques (Optimization)}

The \textit{Optimization} module receives the Ising model or fitness function from the \textit{Evaluation} module and applies a user-selected solver to seek better solutions. 
While \tool supports multiple optimization techniques, its primary focus is on the CIM solver, which simulates the physical dynamics of a network of optical parametric oscillators to fine low-energy states of the Ising Hamiltonian. 
In addition to CIM, \tool also provides classical baseline solvers, including a Brute-Force solver for exhaustive enumeration on small problem instances and classical heuristic algorithms such as Genetic Algorithms (GA)~\cite{lambora2019genetic} and Simulated Annealing (SA)~\cite{delahaye2018simulated}. 

Each solver outputs candidate solutions represented by \textit{Solution Encoding} module, which are subsequently evaluated by the \textit{Evaluation} module.
For Ising-based solvers, the solution is represented as a spin configuration, i.e., a vector of $\pm1$ values over all spin variables. 
For classical solvers, the solution could be represented as a binary vector indicating the selection of test cases. 
The modular design of \tool ensures extensibility, allowing new solvers to be integrated by adhering to the common input and output interface.

The CIM solver is the key optimization component in \tool. 
It is implemented as a software simulation of a measurement-feedback CIM following the Gaussian Approximated Positive-P (GAPP) model proposed by Inui et al.~\cite{inui2022control}. 
In this framework, each spin variable in the Ising model is represented by an optical parametric oscillator whose amplitude evolves according to nonlinear dynamical equations. 
Through iterative measurement and feedback processes, the coupled oscillator network naturally evolves toward low-energy configurations of the Ising Hamiltonian.

During the simulation, the system dynamics iteratively update the oscillator amplitudes while incorporating coupling interactions derived from the Ising coefficients. 
As the evolution progresses, the oscillator states converge to stable configurations, and the signs of their amplitudes correspond to the final spin assignments of the optimization problem. 
This process searches for low-energy states of the Ising model, where minimizing the energy corresponds to obtaining better solutions to the optimization problem.

To support experimentation, the CIM solver exposes several configurable parameters that control the behavior of the dynamical system, as shown in Table~\ref{tab:cim_params}. 
These parameters regulate the nonlinear saturation of oscillators, feedback coupling strength, amplitude stabilization, stochastic noise injection, and numerical integration settings for the simulation.

\begin{table}
	\caption{Configurable Parameters of the CIM Solver}
	\label{tab:cim_params}
	\centering
	\begin{tabular}{p{2cm}p{5.5cm}p{1.5cm}}
		\hline
		\textbf{Parameter} & \textbf{Description} & \textbf{Default} \\
		\hline
		$g^2$ & Control each oscillator's saturation behavior. & $10^{-3}$ \\
		\hline
		$j$ & Control the trade-off between feedback coupling strength and coupling-induced dissipation. & $2$ \\
		\hline
		$\beta$ & Control the strength of amplitude homogeneity correction across spins. & $10$ \\
		\hline
		noise\_scale & Control the magnitude of the stochastic noise term $W_r$. & $1.0$ \\
		\hline
		steps & Control the total evolution time of the CIM dynamics.  & $1000$ \\
		\hline
		$\Delta t$ & Control the integration step size of the CIM dynamics simulation. & $2*10^{-3}$ \\
		\hline
	\end{tabular}
\end{table}

\subsection{Benchmarks}

To provide a platform for assessing Ising-based testing approaches and enabling comparisons with existing methods, \benchmark comprises two reference resources: (i) \textit{Problem Benchmarks}, a collection of existing benchmarks for evaluating testing optimization approaches (e.g., PaintControl and IOF/ROL~\footnote{See \url{https://bitbucket.org/HelgeS/atcs-data/src/master/}}, GSDTSR~\cite{elbaum2014google}), and (ii) \textit{Result Benchmarks}, which contain reference results sourced from prior literature~\cite{wang2024quantum, wang2024test}.
Both resources can be aggregated and extended to incorporate new methods and additional datasets.
Together, they enable users (especially researchers) to develop and evaluate new solvers with \tool and conduct experiments on \benchmark. 
As a result, \tool and \benchmark serve not only as a ready-to-use optimization tool but also as an extensible research platform supporting new problem formulations and solver backends contributed by the community.

\subsection{Result Analysis}

To facilitate the analysis of optimization processes, we implement module \textit{Result Analysis} for examining optimization dynamics with raw results generated by \tool, i.e., \textit{results\textsubscript{r}}.
For instance, we provide convergence curves to show the evolution of fitness scores or spin amplitudes, enabling users to inspect convergence patterns and compare optimization dynamics across various solver configurations.

In addition, the \textit{Result Analysis} module supports benchmarking by providing performance comparison plots that summarize fitness scores across multiple runs using box plots, enabling comparison of performance of different solvers on a given dataset.
The comparisons can be conducted among solvers integrated in \tool or against reference results (i.e., \textit{results\textsubscript{e}}) from \textit{Result Benchmarks}.

\section{Working with \tool and \benchmark}\label{sec: application}

This section presents a step-by-step example showing how to use and extend \tool on \benchmark.
See more detail in our repository: \link.

\subsection{Using \tool and \benchmark}

\paragraph{Preparation}

Solving testing optimization problems requires a dataset representing the problem, along with solution encoding, evaluation strategies, and optimization techniques. 
For TCS and TCM, \tool accepts a CSV file as input to encode the problem, where rows correspond to individual test cases and columns represent test case properties.
An example input from the PaintControl benchmark is shown below.
Each row represents measurable attributes, i.e., execution time (\texttt{time}) and failure rate (\texttt{rate}) of a test case.  
\begin{lstlisting}
,time,rate
0,39050.0,0.13383838383838384
1,1000.0,0.09620253164556962
...
\end{lstlisting}

For integrated problems, solution encoding, evaluation strategies, and optimization techniques are predefined, allowing users to directly configure them in the next step. 
Regarding new problems, users need to provide the dataset together with corresponding solution encoding and evaluation strategies.

\paragraph{Execution}
To ease the use, \tool can be executed using either CLI flags or a YAML configuration file.
An example of using the tool with CLI flags is shown below.

\begin{lstlisting}[language=bash, basicstyle=\ttfamily\footnotesize]
python ising_bench.py test \
	--problem WAOr \
	--library paintcontrol \
	--problem-param \
		effectiveness=['rate'] \
		cost=['time'] \
		minimization=true \
	--solver CIM \
	--solver GA \
	--save-path ./results \
	--convergence-curve \
		spins_amplitude \
		fitness_value
\end{lstlisting}

In this example, the \texttt{WAO\textsubscript{R}} optimization strategy is applied to the \texttt{PaintControl} benchmark. 
The problem parameters specify \texttt{rate} as the effectiveness attribute and \texttt{time} as the cost attribute for the TCM problem (\texttt{minimization=true}). 
(\texttt{CIM} and \texttt{GA}) are executed, with results stored in the \texttt{./results} directory and convergence curves recorded for spin amplitudes and fitness values.

\begin{figure}[h]
	\centering
	\includegraphics[width=0.6\linewidth]{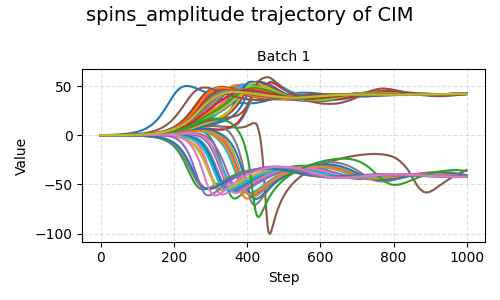}
	\caption{Spin amplitude trajectories produced by the CIM solver in one run (e.g., Batch 1). Each line is the trajectory of one Ising spin, and the y-axis shows the spin amplitude.}
	\label{fig:cim_spin_amplitude}
\end{figure}

\paragraph{Outputs}
Upon completion, \tool decodes the spin configuration and binary vector returned by each solver and outputs the TCS results together with their fitness scores as a JSON file. 
Fig.~\ref{fig:cim_spin_amplitude} is an example of the convergence curves produced by the CIM solver. 
Additional analysis results, such as performance comparison plots, are saved to the configured output directory, enabling direct visual comparison across solvers.
More information can be found in the repository.

\subsection{Extending \tool}

\tool is designed to support extensibility in both problem formulations and solver implementations.

\paragraph{Customizing Problems}
A new problem can be formulated by extending 
the \texttt{BaseProblem} class and registering it with the \texttt{@register\_problem(name)} annotation.
The implementation must define three methods: \texttt{\_calc\_ising()} converting the input data into an Ising model; \texttt{fitness\_function()} evaluating solution quality with respect to the optimization objective; and \texttt{classical\_info()} providing information required by classical solvers, including the number of bits, the optimization direction, and a constraint function.
Once registered, the custom encoding is immediately accessible via configuration.

\paragraph{Customizing Solvers}
A new optimization technique can be integrated by inheriting from the \texttt{BaseIsingSolver} or \texttt{BaseClassicalSolver} class and registering it with the \texttt{@register\_solver(name)} annotation. 
Method \texttt{\_run()} must be implemented to perform solution encoding, for example by returning a spin configuration vector $\mathbf{s} \in \{-1, +1\}^N$ or a binary solution vector.

This plugin-style architecture makes \tool easily extensible, facilitating the integration of new solvers and problem-specific encoding strategies as they emerge.

\section{Discussions}\label{sec:discussions}
\tool and \benchmark serve as a foundation for researchers interested in quantum and quantum-inspired optimization, as they provide both baseline implementations and benchmarks for evaluating and comparing new approaches, which are also extensible. 

Despite the promising design, \tool and \benchmark have several limitations that should be acknowledged. 
First, \tool has not yet been thoroughly evaluated in terms of its applicability, usability, and scalability. For instance, large-scale empirical studies should be conducted on comprehensive benchmark suites or real-world industrial cases to evaluate \tool. 
Second, while the three optimization strategies (Section~\ref{subsec: optStrategies}) integrated in \tool are grounded in established test optimization formulations, their effectiveness across diverse project types, scales, etc. has not yet been systematically evaluated. 
In future work, we plan to incorporate additional optimization strategies into \tool and conduct empirical studies to assess their effectiveness and to compare \tool with existing test optimization solutions. 
We also plan to conduct empirical studies to evaluate CIM’s performance alongside quantum devices. We believe it is important to investigate, for a given software engineering optimization problem, which solver performs best in terms of solution quality, computational efficiency, and scalability.

Along the way, we will also update \benchmark. 
Third, the current interaction model of \tool is command-line based, which may present a usability barrier for practitioners unfamiliar with CLI workflows and limit the tool’s accessibility to a broader engineering audience. In future work, \tool could be integrated with widely used CI/CD platforms, such as GitHub Actions, enabling its adoption in automated software engineering workflows.

\section{Conclusion}\label{sec:conclusion}
This paper presents \tool and \benchmark, a Python-based command-line platform that bridges the gap between test optimization and quantum-inspired algorithms by providing a unified pipeline for solving test optimization problems formulated as Ising models. Though \tool and \benchmark are currently at the prototype stage, its architecture is designed to support extensibility, and its implementation in Python enhances accessibility for researchers and practitioners.


\bibliographystyle{ACM-Reference-Format} 

%


\newpage

\end{document}